\documentclass[pra,twocolumn,showpacs,longbibliography,10pt,aps]{revtex4-1}

\usepackage{array}
\usepackage{graphicx}
\usepackage{verbatim}
\usepackage{color} 
\usepackage{amsfonts}
\usepackage{amsmath}
\usepackage{fancyhdr}

\newcommand{\ket}[1]{\left\vert{#1}\right\rangle}
\newcommand{\bra}[1]{\left\langle{#1}\right\vert}

\begin{document}

\title{Gauge Color Codes in Two Dimensions}
\author{Cody Jones}
\email{ncjones@hrl.com}
\author{Peter Brooks}
\email{pbbrooks@hrl.com}
\author{Jim Harrington}
\email{jwharrington@hrl.com}
\affiliation{HRL Laboratories, LLC, 3011 Malibu Canyon Road, Malibu, CA 90265, USA}

\begin{abstract}
We present a family of quantum error-correcting codes that support a universal set of transversal logic gates using only local operations on a two-dimensional array of physical qubits.  The construction is a subsystem version of color codes where gauge fixing through local measurements dynamically determines which gates are transversal.  Although the operations are local, the underlying code is not topological in structure, which is how the construction circumvents no-go constraints imposed by the Bravyi-K\"onig and Pastawski-Yoshida theorems.  We provide strong evidence that the encoding has no error threshold in the conventional sense, though it is still possible to have logical gates with error probability much lower than that of physical gates.
\end{abstract}


\maketitle

\section{Introduction}
Practical quantum computing requires methods that suppress errors in faulty experimental devices.  Quantum error correction is a broad class of techniques that encode ``logical'' qubits and gates in a subspace of the Hilbert space formed by many more ``physical'' qubits and gates.  The structure of a quantum code has a profound influence on how logical gates are enacted, and hence the total size and execution time of a quantum computation.  This work presents a family of codes that enable a universal set of logic gates with suppression of localized errors to arbitrary order (\emph{i.e.} arbitrary code distance), which we call ``2D gauge color codes.''  This construction is local in two dimensions, meaning that all physical qubits can be arranged on a planar lattice such that physical gates have bounded range independent of code size or distance.  

Early results in quantum error correction established that arbitrarily complex quantum computations can be realized using imperfect physical resources~\cite{Shor1995,Steane1996,Preskill1998}.  Research effort then shifted to finding error-correction schemes that are amenable to practical implementation, such as tolerating high error rates, requiring only short-range connectivity in two or three dimensions, and completing universal logic~\cite{Knill2005,Bravyi2005,Raussendorf2007,Cross2009,Landahl2011,Fowler2012}.  Universal logic, the ability to implement any unitary transform by composing available gates, is essential for quantum computing, but realizing universality has also proven challenging for error correction.  The Eastin-Knill theorem establishes that no stabilizer code can simultaneously implement a universal set of encoded gates transversally~\cite{Chen2008,Eastin2009}.  An encoded logical gate is transversal if it can be implemented by applying independent gates to each physical qubit in the code block.  The individual physical gates may be different; for example, in some families of color codes, a transversal gate is implemented through applying either some gate $U$ or $U^{\dag}$ at alternating sites on the underlying qubit lattice~\cite{Bombin2015,Brown2015,Kubica2015}. 

There are two prominent ways to circumvent the Eastin-Knill theorem and implement fault-tolerant universal logic.  The first is to use teleportation to move quantum information between different codes that provide complementary transversal gates~\cite{Gottesman1999,Zhou2000,Knill2005}.  This category includes the widely studied technique of magic-state distillation~\cite{Bravyi2005,Meier2012,Bravyi2012,Jones2013}.  The substantial resource overhead associated with distillation motivates research into alternative means of achieving universality~\cite{Fowler2013,Jones2013_thesis}.  The second approach is to deform the encoding of quantum information so that it is still protected from localized errors, but the set of allowed transversal gates changes.  These encoding schemes are known as subsystem codes, and their properties were originally studied in the context of quantum memory~\cite{Kribs2005,Poulin2005,Bacon2006,Aliferis2007}.  Recent work has established families of subsystem color codes (also known as ``gauge color codes'') that support a universal set of gates, using ``gauge-fixing'' operations that are local in three dimensions~\cite{Paetznick2013,Bombin2015,Anderson2014,Brown2015,Kubica2015}.

In this paper, we present a new family of gauge color codes that support universal logic but require only two-dimensional locality of operations.  This result would seem to contradict recent no-go theorems that connect dimensionality of the code and the transversal gates that are possible~\cite{Bravyi2013,Pastawski2014}; specifically, a topological code that supports universal transversal logic through local operations must be implemented in three or more dimensions.  The apparent contradiction does not exist because the codes we introduce are not topological in structure (the stabilizer generators do not have local support on the qubit lattice), so they fall outside the scope of the no-go theorems.  Interestingly, our construction produces codes of arbitrarily large error distance, but logical error rates cannot be suppressed arbitrarily toward zero (both statements are supported by arguments in appendices).  Despite lacking an error threshold, the codes can still suppress error in the encoded state below the physical error rate.  We comment further on these matters in Section~\ref{sec::discussion}.  While preparing this manuscript, we became aware of similar work by Bravyi/Cross~\cite{Bravyi2015} and Jochym-O'Connor/Bartlett~\cite{Jochym2015}.

The paper is structured as follows.  Section~\ref{sec::general_construction} describes a generic procedure for constructing a class of triply-even codes from doubly-even codes using the stabilizer formalism.  We call the resulting codes ``chained triply-even codes'' based on the structure of their stabilizers.  Section~\ref{sec::bilayer_color_codes} describes how to construct ``2D gauge color codes, an implementation of the triply-even construction where all operations local in two dimensions.  Section~\ref{sec::discussion} discusses properties of 2D gauge color codes and connections to other results in the literature.  The appendices contain supporting proofs and an algorithm for syndrome decoding.

\section{A Procedure for Constructing Triply-Even Codes}
\label{sec::general_construction}
\subsection{Preliminaries}
We begin by defining terminology and proving preliminary results.  Throughout the paper, $X$ or $Z$ stand for single-qubit Pauli operators $\sigma^x$ or $\sigma^z$, respectively.  Subscript ``$L$'' denotes logical states or operators (\emph{e.g.} $X_L$ is logical $X$ for an encoded qubit).  For an operator $P$ that is a tensor product of single-qubit physical Pauli operators, define the weight $w(P)$ to be the number of non-identity terms in the tensor product.  A Calderbank-Shor-Steane (CSS) code~\cite{Calderbank1996,Steane1996b} is said to be ``even'' if the weight of $s^x$ is even for every element $s^x$ of the $X$-type stabilizer group $\mathcal{S}^x$, or $w(s^x) \equiv 0 \; (\mathrm{mod} \; 2) \; \forall s^x \in \mathcal{S}^x$.  Similarly, a doubly-even CSS code is identified by the property $w(s^x) \equiv 0 \; (\mathrm{mod} \; 4) \; \forall s^x \in \mathcal{S}^x$.  A doubly-even CSS code has the additional property that transversal $S = Z^{1/2}$ preserves the stabilizer, so it is a logical operation, such as $S_L$ or $S_L^{\dag}$.  This property generalizes for $k$-even codes, as we now demonstrate.  Consider an $[[n,1,d]]$ CSS code where $n$ is odd and $X_L = X^{\otimes n}$ ($X$ applied to all physical qubits).  The logical basis states are
\begin{equation}
\ket{0_L} = \frac{1}{\sqrt{|\mathcal{S}^x|}}\sum_{s^x \in \mathcal{S}^x} s^x\ket{0}^{\otimes n},
\label{logical_zero}
\end{equation}
\begin{equation}
\ket{1_L} = X_L \ket{0_L}.
\end{equation}
Quantity $|\mathcal{S}^x|$ is the number of unique elements of the $X$-type stabilizer group, including identity.  If this code is $k$-even, then every $X$ stabilizer element has $w(s^x) \equiv 0 \; (\mathrm{mod} \; 2^k)$, and it follows that the Hamming weight of every term in the sum Eqn.~(\ref{logical_zero}) is likewise a multiple of $2^k$.  As a result, transversal application of 
\begin{equation}
R_Z(\pi/2^{k-1}) = \ket{0}\bra{0} + \exp(i\pi/2^{k-1})\ket{1}\bra{1}
\end{equation}
acts trivially on $\ket{0_L}$.  By contrast, $w(X^{\otimes n} s^x) \equiv n \; (\mathrm{mod} \; 2^k)$ is the Hamming weight of all terms in the expansion of $\ket{1_L}$.  Applying transversal $R_Z(\pi/2^{k-1})$ to an arbitrary logical qubit implements logical $R_Z(n\pi/2^{k-1})$: 
\begin{align}
R_Z(\pi/2^{k-1})^{\otimes n} & \left( \alpha \ket{0_L} + \beta \ket{1_L}\right) \nonumber \\
= & \alpha \ket{0_L} + \exp(i \pi n/2^{k-1}) \beta \ket{1_L}.
\end{align}
Since $n$ is odd, it follows that logical gates $R_Z(m\pi/2^{k-1})$ for any integer $m$ can be implemented with transversal physical gates. For example, triply-even codes~\cite{Betsumiya2012} have the property that transversal $T = R_Z(\pi/4)$ is a logical operation, and it is some odd power of logical $T$ for codes satisfying the conditions above.  Codes with transversal $T$ gate are desirable for completing universal gate sets in quantum logic since this operation is outside the Clifford group~\cite{Bravyi2005,Knill2005,Raussendorf2007,Bravyi2012,Paetznick2013}.

\subsection{Recursive Construction of Triply-Even Codes}
We describe a procedure to transform a doubly-even code to a triply-even code using code deformation (manipulations of stabilizers).  For notational convenience, we will index a family of codes by nonnegative integer $t$, the minimum number of correctable errors for that code.  Starting with an order-$t$ doubly-even code in an arbitrary logical state, the procedure requires a second doubly-even code block (with same stabilizers) and a triply even code of order $(t-1)$, where the logical qubits of these two additional code blocks are entangled in a logical Bell pair~\citep{Anderson2014}.  The two doubly-even code blocks are fused using gauge fixing~\cite{Paetznick2013,Bombin2015}, which teleports the logical qubit of the original doubly-even code into a triply-even code of order $t$.

Let $D_t$ or $T_t$ denote a doubly-even code or triply-even code of order $t$, respectively, with one logical qubit and distance $d = 2t + 1$.  To be clear on notation, we refer to the $T = R_Z(\pi/4)$ gate, which is the transversal operation of interest in the triply-even code, without a subscript.  To make our construction work, we place some additional constraints on code properties.  First, $D_t$ or $T_t$ is an $[[n,1,2t+1]]$ code where $n$ is odd and where transversal $X$ is logical $X$ (\emph{i.e.} $X^{\otimes n} = X_L$).  Second, for both $D_t$ and $T_t$, the code size is related to distance by
\begin{align}
n & \equiv 1 \; (\mathrm{mod} \; 8), \; \mathrm{for} \; t \; \mathrm{even} \nonumber \\
n & \equiv 7 \; (\mathrm{mod} \; 8), \; \mathrm{for} \; t \; \mathrm{odd}
\label{eqn::code_size_constraints}
\end{align}
The two-step conversion procedure requires a second copy of the code $D_t$; to distinguish the two doubly-even blocks in usage, let $D_t^{(a)}$ denote the initial code block with arbitrary logical qubit (hence superscript ``a'').  First, prepare the logical qubits of additional code blocks $T_{t-1}$ and $D_t^{(b)}$ in an entangled Bell state (hence ``b''), and a method for doing so is described in Section~\ref{sec::bilayer_color_codes}.  Second, perform weight-four $Z$ measurements on every pair of qubits in $D_t^{(a)}$ and the matching pair of qubits in $D_t^{(b)}$ in the same locations.  For the purposes of this paper, we say that these $Z$-type measurements ``fuse'' the two codes, because the matching generators of the $X$ stabilizers for these codes become welded together.  The conversion process depends recursively on a triply-even code of lower order, as shown in the following compact form:
\begin{equation}
\overbrace{T_{t-1} \vert D_t^{(b)}}^{\mathrm{Bell}} \underset{ZZ}{\Longleftrightarrow} D_t^{(a)} \rightarrow T_t.
\label{eqn::recursion}
\end{equation}
The two notations we introduce are $F \vert G$ with label ``Bell'', which represents forming a logical Bell state between code blocks $F$ and $G$, and symbol $\underset{ZZ}{\Longleftrightarrow}$, which denotes fusing the two identical codes immediately to the left and right of the operator (using the set of weight-four $Z$ measurements).  After conversion, the logical qubit of $D_t^{(a)}$ becomes that of $T_t$.  In Appendix~\ref{app::properties}, we prove that $T_t$ can correct any configuration of $t$ or fewer independent physical errors.

The procedure described above promotes a doubly-even code to a triply-even code where transversal operation $T^{\otimes n}$ is either logical $T$ or $T^{\dag}$, depending on order $t$; by construction, the weight of any logical $X$ operator in our triply-even codes follows Eqn.~(\ref{eqn::code_size_constraints}).  As a result, a triply-even code of any order can be generated using a family of doubly-even codes that extend to any order, starting the recursion chain from $D_0 = T_0$ (a bare physical qubit), such as:
\begin{equation}
\overbrace{D_0 \vert D_1}^{\mathrm{Bell}} \underset{ZZ}{\Longleftrightarrow} \overbrace{D_1 \vert D_2}^{\mathrm{Bell}} \underset{ZZ}{\Longleftrightarrow} D_2^{(a)} \rightarrow T_2.
\label{eqn::recursion_example}
\end{equation}
The arbitrary logical qubit of $D_2^{(a)}$ becomes the logical qubit of $T_2$.  One way to understand the code conversion is that $D_t^{(a)}$ and $T_t$ are specific configurations of the same subsystem code, and one changes the configuration through ``gauge fixing''~\cite{Kribs2005,Poulin2005,Bacon2006,Paetznick2013,Bombin2015,Anderson2014}.  The procedure in Eqn.~(\ref{eqn::recursion_example}) combines doubly-even codes using alternating links of logical Bell states and bilayer fusion, so we call the result a ``chained triply-even code.''  The distinction is needed since there are triply-even codes not described by this procedure, such as the concatenated 15-qubit Reed-Muller code~\cite{Paetznick2013,Anderson2014}.  Appendix~\ref{app::properties} proves that an order-$t$ chained triply-even code has transversal $T$ gate and can correct $t$ errors, for any $t$.  Despite extending to arbitrary order, Appendix~\ref{app::properties} also argues that chained triply-even codes do not have an error threshold in the conventional sense.

\section{2D Gauge Color Codes}
\label{sec::bilayer_color_codes}
Two-dimensional color codes provide a suitable family of doubly-even codes to construct chained triply-even codes to arbitrary order.  Moreover, these codes support the formation of logical Bell pairs using local measurements through ``lattice surgery''~\cite{Horsman2012,Landahl2014}.  We focus on the triangular [4.8.8] ``squares and octagons'' codes~\cite{Landahl2011,Stephens2014,Delfosse2014,Kubica2015}, though the same techniques can be applied to triangular [6.6.6] ``hexagons'' color codes~\cite{Landahl2011,Bravyi2015,Jochym2015} and any other color codes that are doubly-even in a generalized sense~\cite{Bombin2015,Kubica2015}.  For the remainder of the paper, we use $D_t$ to specifically refer to the triangular [4.8.8] color code of order $t$.  When qubits comprising the color-code blocks are arranged in a bilayer of planar lattices, all of the necessary operations for gauge fixing are local.  By merging the two planes, the construction is still local in a two-dimensional lattice of qubits.

\begin{figure}
	\centering
  \includegraphics[width=8.3cm]{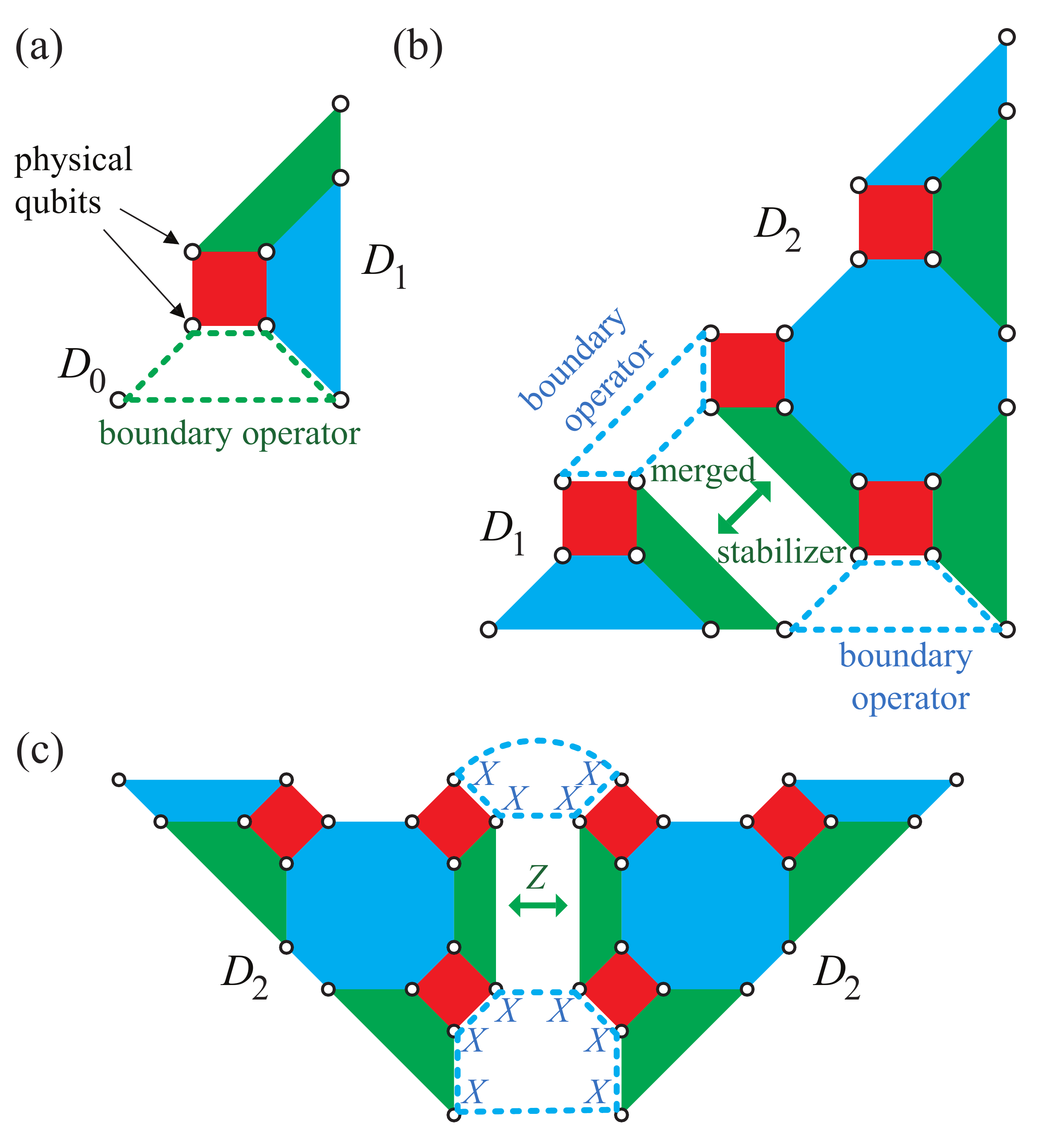}
  \caption{(Color) Boundary measurement between color codes. (a)~Codes $D_0$ and $D_1$ are projected into logical Bell state by measuring weight-four $X$ and $Z$ operator shown in dashed lines.  (b)~Codes $D_1$ and $D_2$ are entangled through two boundary operators.  Green check operators along the boundary are merged together, as indicated by the two-sided arrow.  (c)~Logical $X_L X_L$ between two $D_2$ color codes (note $X$'s inside boundary operators) will only merge green check operators of $Z$ type.}
  \label{boundary_measurement}
\end{figure}

\begin{figure*}
	\centering
  \includegraphics[width=\textwidth]{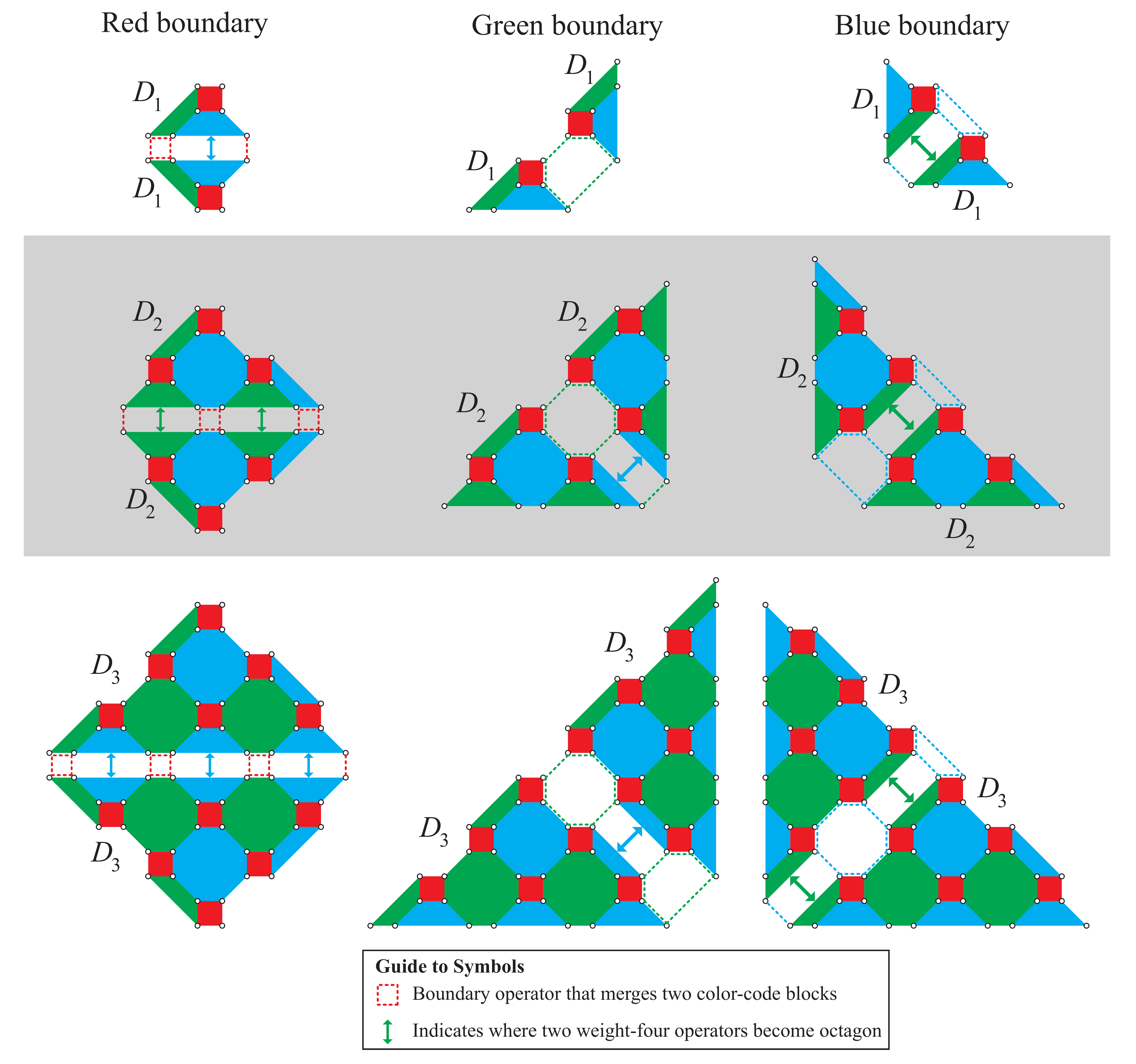}
  \caption{(Color) Examples of boundary measurement between triangular color codes.  The grid of examples shows orders $t = 1, 2, 3$ by row and red/green/blue boundaries by column.  Larger codes follow these patterns, with squares or octagons (depending on color) as operators in the center of the boundary, and one weight-two or weight-six operator at an endpoint of the boundary.}
  \label{fig::boundary_same_d}
\end{figure*}

\subsection{Boundary and Bilayer Gauge Operations}
For triangular color codes, the boundary measurement procedure is a two-step process of merging and separating two code blocks using lattice surgery~\cite{Horsman2012,Landahl2014}.  In the first step, merge two codes into one color code fabric, using the translational symmetry of the ``square'' and ``octagon'' operators.  The merging is implemented by changing the local measurements along the common boundary of the codes, either by inserting new operators or modifying existing ones.  Since the boundary operators follow the tiling pattern of the color code, the weight of each is an even number between two and eight.  We provide several examples in Fig.~\ref{boundary_measurement}.  When only $X_L X_L$ is required, the existing $X$-type check operators of the two color codes are unchanged; new $X$-type check operators are measured, and some $Z$-type check operators along the boundary are merged together (the situation is reversed for $Z_L Z_L$).  To make a projective Bell-basis measurement (both $X_L X_L$ and $Z_L Z_L$), the procedure inserts new $X$- and $Z$-type check operators along the boundary and modifies existing check operators of both types.   The continuous fabric of overlapping check operators can detect localized errors that occur during this procedure.

The second step is to separate the color codes by measuring their original stabilizers.  After merging and separating two color codes, the logical qubits have been jointly measured ($X_L X_L$, $Z_L Z_L$, or both).  The result of this logical measurement is obtained by taking the joint parity of boundary-operator measurements.  Moreover, the protocol provides detection of errors up to the lesser distance of the two participating color codes.  When measurements are faulty, check operators must be measured multiple times and the results jointly processed, as in other fault-tolerant QEC schemes~\cite{Steane1996,Raussendorf2007,Cross2009,Fowler2012,Stephens2014,Stephens2014b}.

Clearly, the boundary-measurement procedure requires that the two color codes be of similar size.  For our purposes, we require boundary measurements between two color codes of the same order (used for logical CNOT) or two codes with order differing by one (used to make logical Bell states for the double-to-triple code conversion in Section~\ref{sec::general_construction}), and it happens that both exist for the [4.8.8] triangular color codes of any size, as we now demonstrate.  

\begin{figure*}
	\centering
  \includegraphics[width=\textwidth]{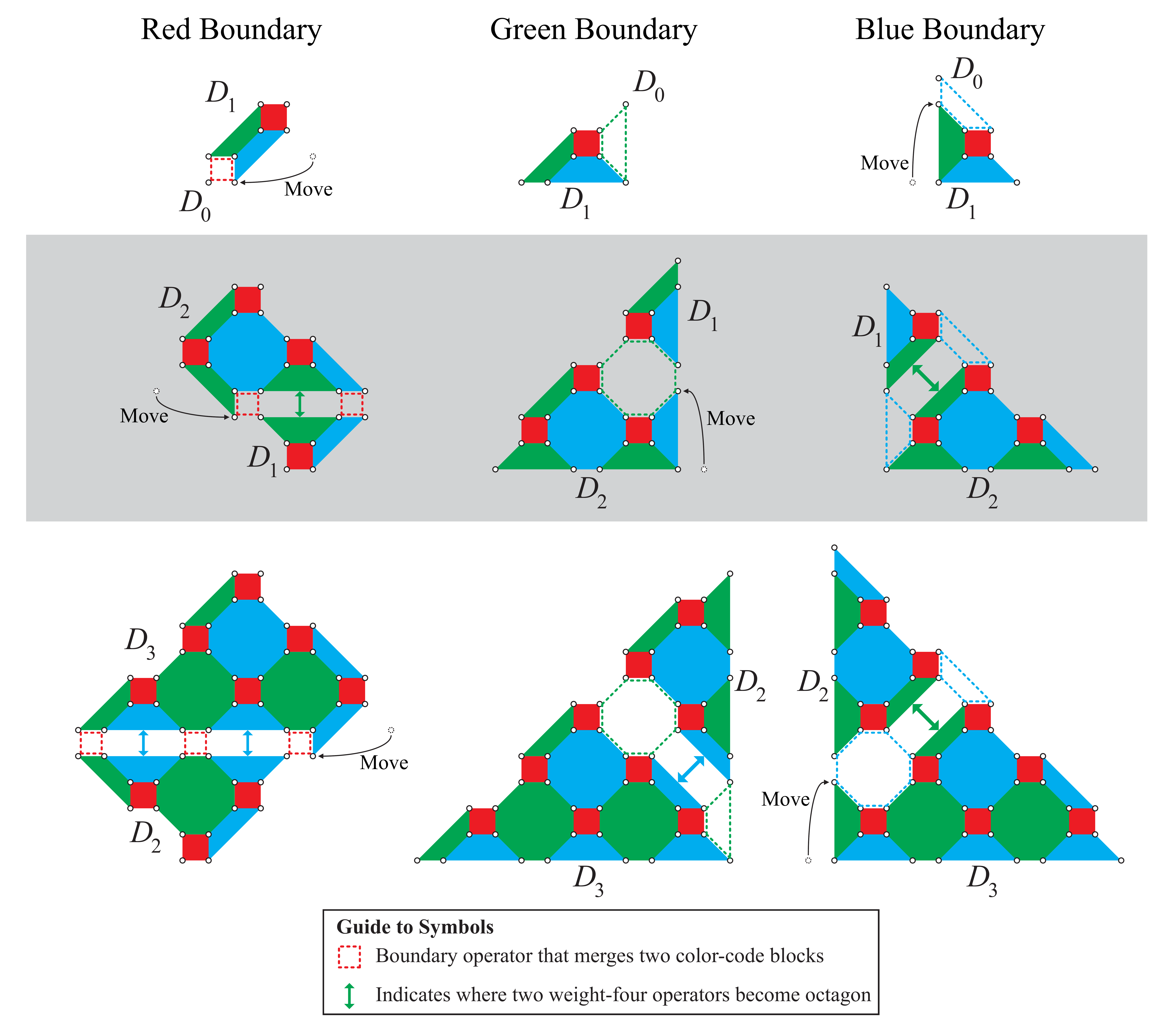}
  \caption{(Color) Examples of boundary measurement between color codes of orders $(t-1)$ and $t$.  The grid shows orders $t = 1,2,3$ by row and red/green/blue boundaries by column.}
  \label{boundary_different_d}
\end{figure*}

We choose to implement boundary measurements in a manner that continues the tiling pattern of the color codes.  This approach proves sufficient and simplifies our analysis, though it is likely that other forms of boundary operations are possible.  Figure~\ref{fig::boundary_same_d} shows multiple panels depicting boundary measurement along the three edges of a given triangular color code for order $t = 1, 2, 3$.  For larger codes, the pattern at the boundary is a continuous ``squares and octagons'' fabric in the center, with special treatment at the corners of the triangular codes (shown in Fig.~\ref{fig::boundary_same_d}), which are endpoints of the boundary.  Note that the codes are mirrored along the boundary.  Each of three different boundaries for a triangular code is associated with a color.  The boundary operators (denoted with dashed lines) are of this color, and the joint parity of their measurement values is the $X_L X_L$ (or $Z_L Z_L$) measurement value.  The merged stabilizers (denoted with arrows) can be used for error detection.  Note that one endpoint of the boundary will have a weight-two or weight-six operator.  Individual triangular color codes have stabilizer checks of weight four or eight, but these new operators correspond to logical operators.  Since these boundary operators are still even in weight, $X_L X_L$ and $Z_L Z_L$ boundary measurements commute, as expected, and they could be measured concurrently.

\begin{figure}
	\centering
 \includegraphics[width=8.3cm]{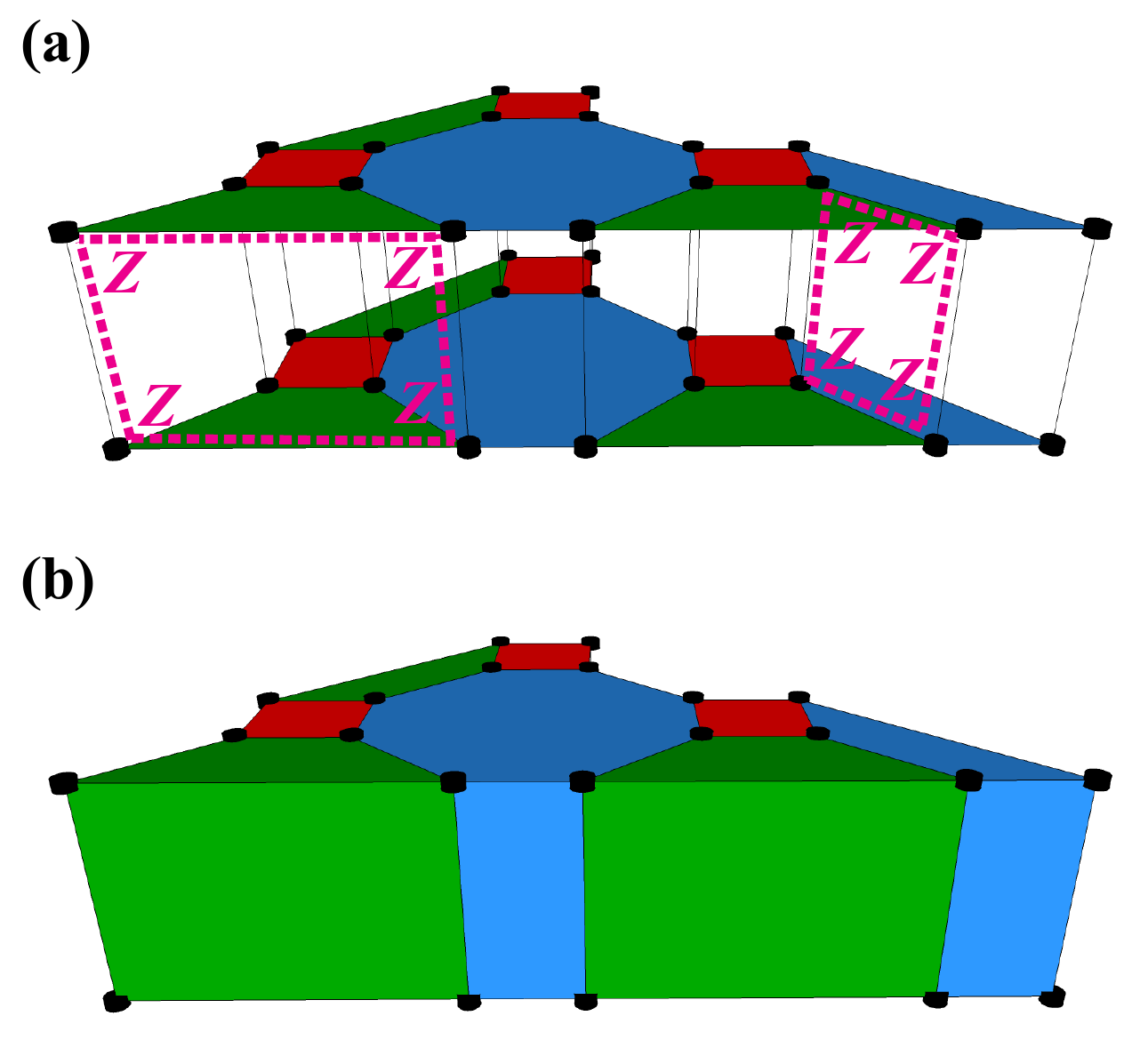}
  \caption{(Color) Example of fusion in bilayer of two $D_2$ code blocks, using [4.8.8] triangular color codes~\cite{Landahl2011}.  (a)~Bilayer stack of two code blocks, showing two examples of the weight-four $Z$ gauge operators that are measured.  Our proposed approach is to measure weight-four $Z$ operators for every matching pairs of edges in the [4.8.8] lattices.  (b) The $X$ stabilizers of the two code blocks are fused by the gauge fixing operation.}
  \label{fig::fused_bilayer}
\end{figure}

Boundary measurement between color codes of orders $(t-1)$ and $t$ is similar to that between codes of the same order, but there are some slight modifications.  Figure~\ref{boundary_different_d} shows several examples.  As before, the logical $X_L X_L$ or $Z_L Z_L$ operator is the joint parity of boundary operators of a specified color.  One difference from Fig.~\ref{fig::boundary_same_d} is that only some of the blue and green boundaries align naturally; specifically, the larger code must have a ``concave edge'' near a corner of the triangular outline, drawn as the bottom-right corner in Fig.~\ref{boundary_different_d}.  For the other cases, alignment is possible by moving the position of one corner qubit, but the code is still local in two dimensions.  

By using [4.8.8] triangular color codes~\cite{Landahl2011} as the doubly-even codes in our construction, the fusion process (measuring weight-four $Z$ operators between two copies of $D_t$) is local with two planar codes stacked in a bilayer structure.  The general prescription from Section~\ref{sec::general_construction} to measure all matching pairs of $Z$ operators between the two doubly-even code blocks requires $2t^4 + O(t^3)$ measurements, which is both over-complete and non-local.  Instead, we propose to only measure pairs of qubits that are connected by edges in the [4.8.8] lattice (in other words, the edges at the boundary of any stabilizer generator for the doubly-even code), as depicted in Fig.~\ref{fig::fused_bilayer}.  This approach requires $3t^2 + O(t)$ measurements, which is over-complete by exactly a factor of three, but all measurements are local in the bilayer.  Moreover the factor-of-three redundancy in gauge measurements is very similar to the proposal for single-shot error correction in 3D gauge color codes~\cite{Bombin2014}, which could have significant implications for fault tolerance by detecting gauge-measurement errors (though we do not analyze the matter here).

\begin{figure}
	\centering
  \includegraphics[width=8.3cm]{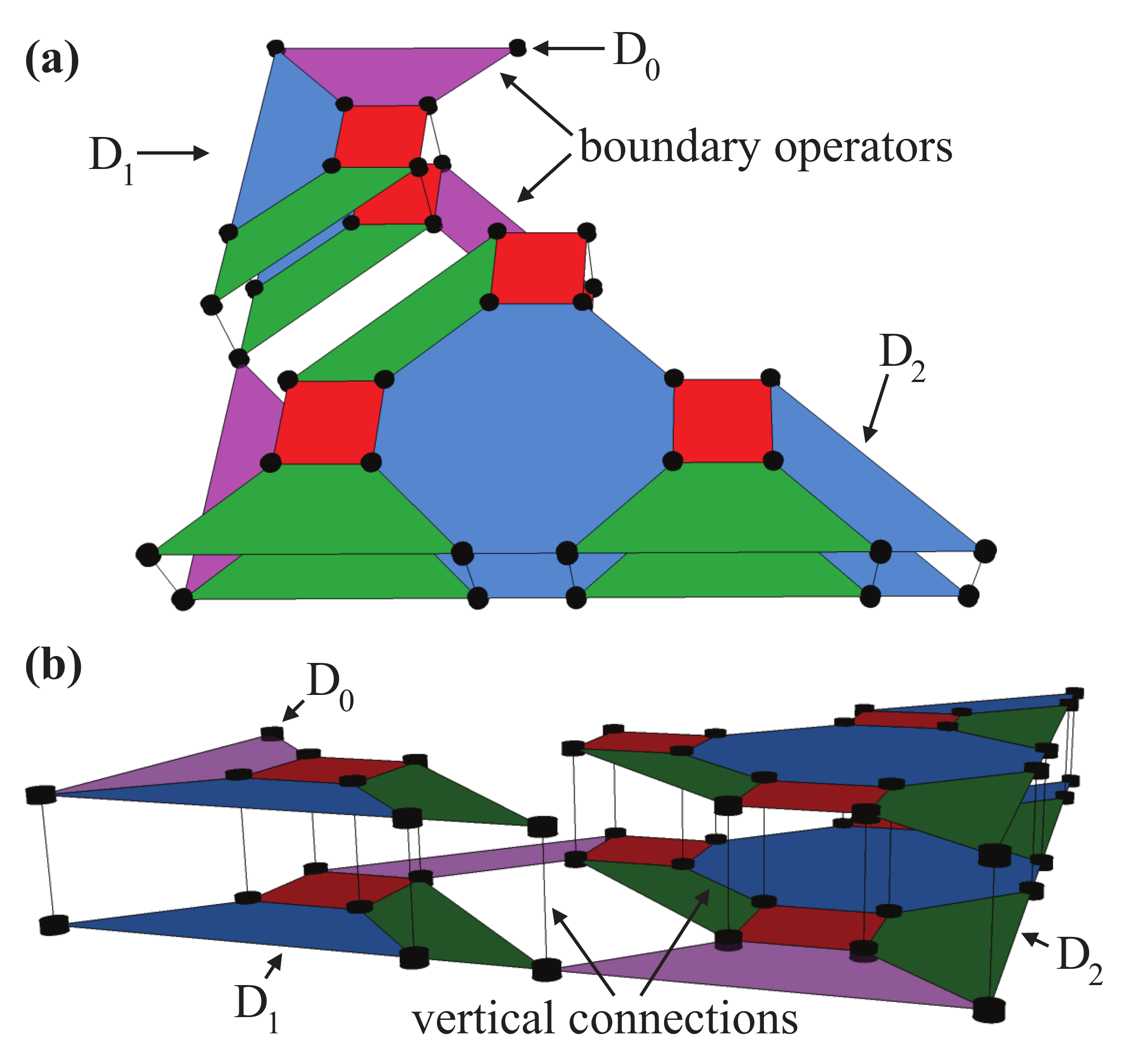}
  \caption{(Color) 2D gauge color code $T_2$ in a bilayer structure.  (a)~Top view, showing the lateral connections through boundary operators (purple).  (b)~Side view, showing the bilayer structure with vertical connections as black lines.  Each weight-four $Z$ measurement in fusion consists of two neighboring pairs of qubits connected by these vertical lines.  Notice that the boundary operators alternate between top and bottom layer, following the recursion in Eqn.~(\ref{eqn::recursion}).}
  \label{fig::bilayer_structure}
\end{figure}

Putting all of the elements together, Fig.~\ref{fig::bilayer_structure} shows how to construct $T_2$ in a bilayer structure.  The ability to create logical Bell states along any of the three boundaries of [4.8.8] color codes provides significant freedom for placement of triangular color codes in a two-dimensional array of qubits.  The construction can be embedded in two-dimensions by, for example, placing two qubits at every site in the [4.8.8] lattice, and all gauge operators remain local.

\subsection{Universal Set of Encoded Gates}
\begin{figure}
	\centering
  \includegraphics[width=8.3cm]{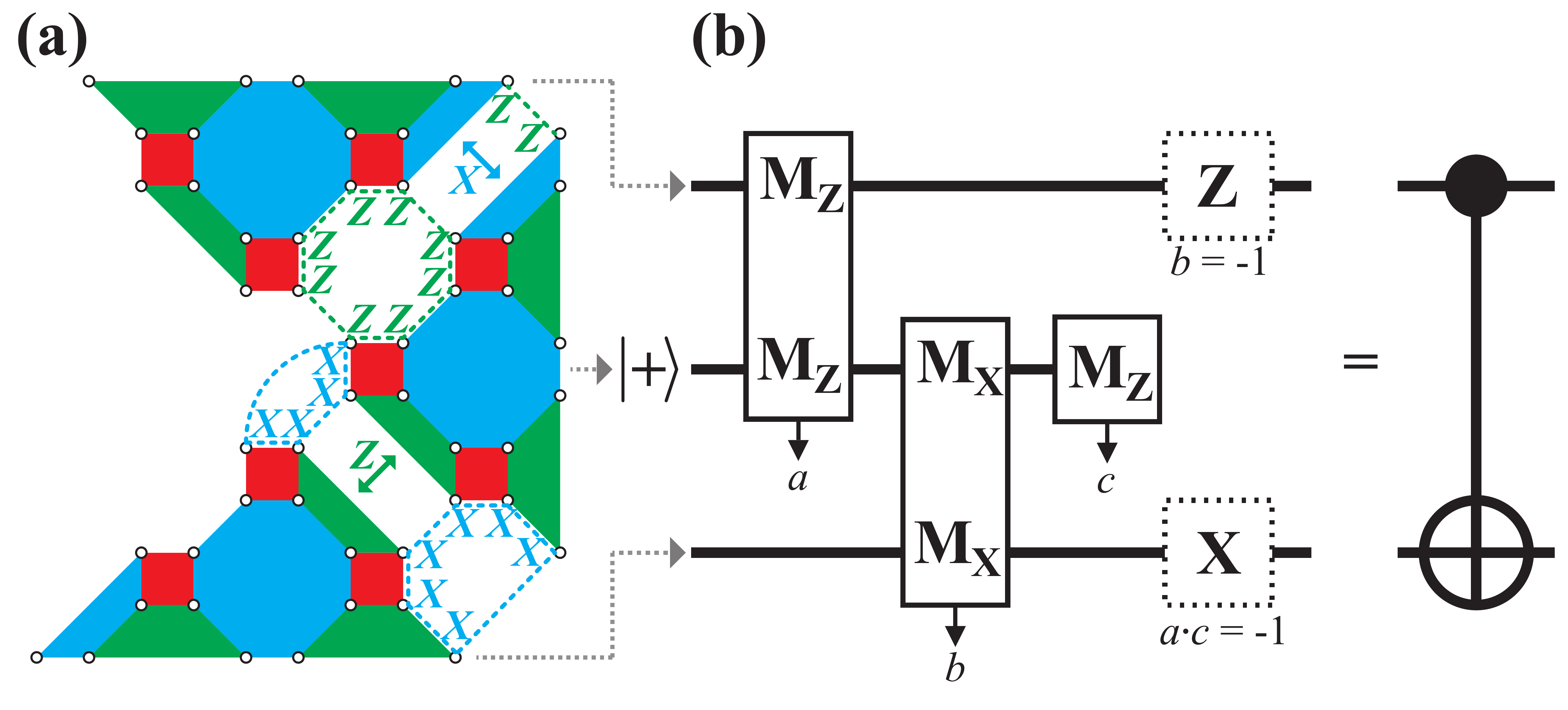}
  \caption{(Color) Logical CNOT using single-qubit initialization/measurement and boundary measurement between color codes.  (a)~Measuring the dashed-line operators at a boundary labeled with $Z$'s ($X$'s) projectively measures the operator $Z_L Z_L$ ($X_L X_L$), as indicated in the circuit at right.  $X$ stabilizers along the boundary are fused together during $Z_L Z_L$ measurement, and vice versa, as indicated by the two-sided arrows.  The middle color code is the ancilla logical qubit.  (b)~Circuit for logical CNOT through parity measurements.  Logical Pauli corrections shown in dashed boxes are conditioned on the measurement results $(a,b,c)$, which are each $\pm1$.}
  \label{color_code_CNOT}
\end{figure}

Through the gauge fixing described above, 2D gauge color codes support single-qubit logical Clifford+$T$ gates, which is a universal set for one logical qubit.  The addition of a logical CNOT gate completes a universal set for approximating arbitrary logical circuits.  As both the triangular color codes and 2D gauge color codes are CSS codes, CNOT could be implemented transversally between the code blocks, but this would weaken the locality of operations.  We propose instead to implement CNOT using $X_L X_L$ and $Z_L Z_L$ measurements along the boundary of two triangular color codes.  The sequence of operations in Fig.~\ref{color_code_CNOT} shows how to implement logical CNOT between two code blocks with the aid of a logical-ancilla code block using lattice surgery~\cite{Horsman2012,Landahl2014}.  For CSS codes, state initialization in logical $X$ or $Z$ basis is achieved by preparing all physical qubits as $\ket{+}^{\otimes n}$ or $\ket{0}^{\otimes n}$, respectively, then measuring the check operators of the code to project into $\ket{+_L}$ or $\ket{0_L}$, as in Eqn.~(\ref{logical_zero}).  The measurements are used to correct the stabilizer, which is initially projected into a random configuration.  Likewise, logical measurement in $X$ or $Z$ basis is implemented by performing transversal measurements in that basis over the entire code block, then using parity combinations of the results for correcting errors and determining the logical measurement outcome.  

Logical gates in 2D gauge color codes are implemented as follows.  One performs all logical Clifford gates on qubits in triangular color codes using transversal gates for single-qubit operations and boundary measurements for CNOTs.  When a non-Clifford $T$ gate is required, a triangular color code is temporarily promoted to a triply-even code, then the $T$ gate is applied transversally, then the code reverts to a triangular color code.  The last step can be performed by simply measuring in the $X$ basis all physical qubits except those in the highest-order triangular block, which is $D_t^{(a)}$ in Eqn.~(\ref{eqn::recursion}).  The logical qubit is returned to the triangular color code, and the $X$-basis measurement results are combined with subsequent triangular-code stabilizer measurements to detect errors that could have occurred during the transversal $T$ gate.

\begin{figure}
	\centering
  \includegraphics[width=8.3cm]{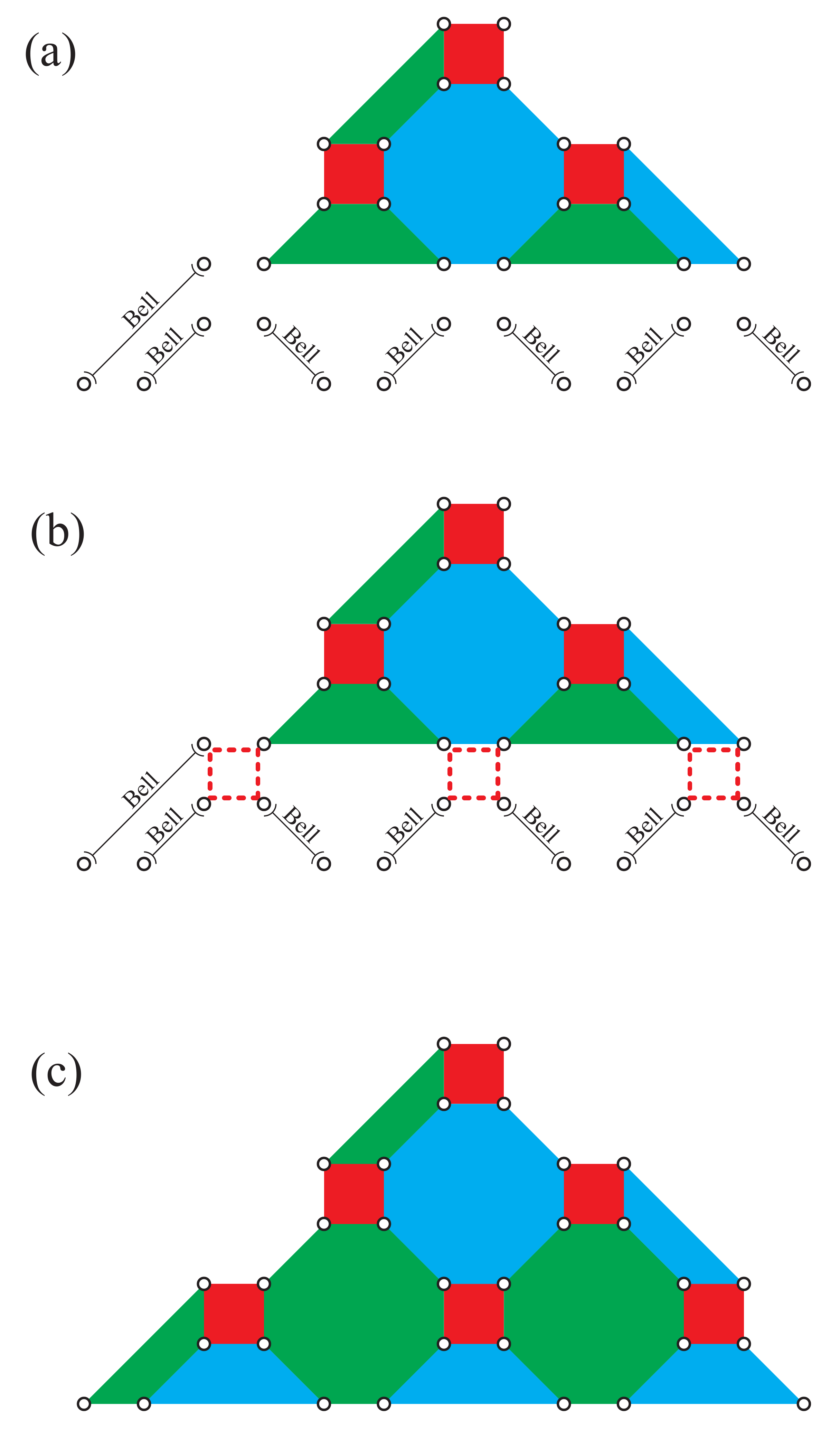}
  \caption{(Color) Code expansion showing how to convert a triangular color code from order two to three.  (a)~The existing order-two code is supplemented with 14 additional physical qubits needed for the expanded code, and they are prepared as Bell pairs along every diagonal edge, as drawn here.  (b)~Measuring the red squares in both $X$ and $Z$ stabilizers will project into the order-three code.  (c)~The remaining blue and green stabilizers are measured for error detection.  The procedure detects errors because the Bell states are stabilized by $XX$ and $ZZ$, and one can form any of the new blue and green stabilizers from products of prior ones along the bottom boundary and weight-two operators from the Bell pairs, enabling error detection.}
  \label{fig::code_expansion}
\end{figure}

As a final consideration, a logical $T$ gate will have a higher probability of logical error than a logical Clifford gate in a triangular color code at the same order, because the $T$ gate requires promotion to a triply-even code with weaker error detection (see Appendix~\ref{app::properties}).  To mitigate this, one could increase the order of the triangular color code with an arbitrary logical state before converting to a 2D gauge color code, so that the transversal $T$ gate is performed in a code with higher error distance than when Clifford gates are performed on a triangular color code.  Figure~\ref{fig::code_expansion} shows how to promote an order-$t$ [4.8.8] triangular color code to order ($t+1$) in a fault-tolerant way.  First, place $4t+6$ additional physical qubits along the border associated with a red logical operator (see Section~\ref{sec::bilayer_color_codes}.  Second, prepare these physical qubits in physical Bell pairs along what will become the ``diagonal'' edges of blue and green stabilizers (as drawn here).  Third, measure the stabilizer checks of the order-$(t+1)$ code, which complete the code expansion.  Specifically, the new ``square'' operators project $D_t$ with additional physical Bell pairs into $D_{t+1}$.  One can explicitly verify that every logical operator of the old code is converted into one of the new code, and the stabilizers have been measured.  The procedure can correct any error configuration of weight $t$ or less during conversion (using the original code), and the code after conversion can correct $t+1$ physical errors.  Moreover, since all operations are local, the order of a triangular code can be increased arbitrarily in constant time, which is essentially the same process as defect movement in topological codes~\cite{Raussendorf2007,Landahl2011,Fowler2012,Fowler2013}.

\section{Discussion}
\label{sec::discussion}
Several recent developments in the research literature led to the constructions developed here.  First, the description of triply-even codes~\cite{Betsumiya2012} provided a target for codes where the $T$ gate is transversal.  Second, the construction of logical Bell states enabled conversion between the doubly-even 7-qubit Steane code ($D_1$) and triply-even 15-qubit Reed-Muller code ($T_1$)~\cite{Anderson2014}.  The gauge fixing technique, proposed for $T_1$ by Paetznick and Reichardt~\cite{Paetznick2013} and generalized to a family of 3D gauge color codes by Bombin and others~\cite{Bombin2015,Anderson2014,Brown2015,Kubica2015}, provided a guiding principle for circumventing the Eastin-Knill theorem.  Finally, the lattice surgery technique, proposed for surface codes~\cite{Horsman2012} and adapted to color codes~\cite{Landahl2014}, provided a local way of measuring high-weight operators, which is necessary to circumvent the Bravyi-K\"onig and Pastawski-Yoshida theorems~\cite{Bravyi2013,Pastawski2014}.  Another connection to recent research is that the non-subsystem implementation of $T_2$ is the 49-qubit triply-even code identified in Ref.~\cite{Bravyi2012}.

A natural question to ask is how 2D gauge color codes compare to other 2D-local codes that complete universality through magic-state distillation, such as surface codes, but an answer requires further analysis like the detailed numerical studies of Refs.~\cite{Cross2009,Fowler2012,Jones2013_thesis,Stephens2014,Stephens2014b}.  Appendix~\ref{app::properties} argues that chained triply-even codes, including 2D gauge color codes, do not have a threshold for suppressing logical error rate arbitrarily.  However, it still appears possible to suppress error rate well below the physical error rate, but a conclusive statement would require further investigation.  While preparing this manuscript, we became aware of two similar proposals for constructing triply-even codes in two dimensions~\cite{Bravyi2015,Jochym2015}.  Both proposals appear to be based on [6.6.6] color codes, and at first glance it does not appear as convenient to measure boundary operators for logical Bell states in these codes.  The authors propose either measuring high-weight stabilizers along the boundary or including strings of ancilla physical Bell pairs.  In both cases, the measurement of logical Bell-state stabilizers will have higher error rate than in our proposal, which may require more measurement repetitions.  Moreover, as our analysis in Appendix~\ref{app::properties} shows that these stabilizers are the weak links in the error detection process, the construction with [4.8.8] color codes may yield lower logical error rates.  In any case, many of the properties that we derive for chained triply-even codes should extend to those constructions (including the syndrome decoder) by making suitable adjustments for the generalized form of doubly-evenness or triply-evenness used therein.  Taken together, these results provide an intriguing new approach to performing universal logical operations with physical operations that are local in just two dimensions.

\begin{acknowledgments}
The authors thank Adam Paetznick and Josh Baron for helpful discussions.  This research was developed with funding from the Defense Advanced Research Projects Agency (DARPA).  The views, opinions, and/or findings contained in this material are those of the authors and should not be interpreted as representing the official views or policies of the Department of Defense or the U.S. Government.
\end{acknowledgments}

\appendix

\section{Properties of Chained Triply-Even Codes}
\label{app::properties}
We prove that the recursive construction of $T_t$ in Eqn.~(\ref{eqn::recursion}) produces a triply-even code of order $t$ in three steps.  First, we show that if the doubly-even codes satisfy Eqn.~(\ref{eqn::code_size_constraints}) and have $X_L = X^{\otimes n}$, then so will the resulting triply-even codes.  Second, we prove that $T_t$ is in fact a triply-even code with transversal $T_L$ by examining its $X$ stabilizer group.  Third, we prove that $T_t$ can correct $t$ errors by examining the logical operators of this code.  Finally, we present an argument that chained triply-even codes do not have an error threshold, in the following sense.  For a fixed physical error rate, there is some lower bound on the achievable logical error rate.  Nevertheless, we also argue that the logical error rate is a strongly nonlinear function of the physical error rate, so it is possible to have logical error rate that is well below the physical error rate.

For pedagogical purposes, we group the $X$ stabilizers affected by code conversion into two types, ``type-B'' and ``type-F.''  The type-B stabilizers originate from the logical Bell pairs, and they have high weight given by $\mathrm{size}(D_{t-1}) + \mathrm{size}(D_t)$.  Type-F stabilizers are isolated by the fusion process between two copies of $D_t$, so their weight is twice that of the generators in $D_t$.  Imagine the two identical code blocks $D_t^{(a)}$ and $D_t^{(b)}$ are arranged in two planes stacked in a bilayer structure, such as the example in Fig.~\ref{fig::fused_bilayer}.  A pair of qubits in the same horizontal location are connected by a ``vertical edge.''  The fusion process takes the two copies of a matching weight-$w$ $X$ stabilizer for these codes and fuses them into a weight-$2w$ type-F stabilizer, also shown by the cells in Fig.~\ref{fig::fused_bilayer}b.  Strictly speaking, the type-F stabilizers are present before fusion, as each is the product of two matching face stabilizers from two copies of $D_t$, but the individual face stabilizers are eliminated by fusion since they do not commute with the weight-four Z operators for fusion (see Fig.~\ref{fig::fused_bilayer}a).

\emph{Requirements on Code Properties} --- We require that the doubly-even codes used in our construction satisfy Eqn.~(\ref{eqn::code_size_constraints}) and have $X_L = X^{\otimes n}$.  Note that these conditions are satisfied by triangular [4.8.8] color codes~\cite{Landahl2011}.  Using the recursive construction in Eqn.~(\ref{eqn::recursion}), the same properties hold for $T_t$, which we now prove inductively.  $T_0$ is a single unencoded qubit which trivially satisfies these conditions.  Because two copies of doubly-even code $D_1$ will require a number of qubits that is $6 \; (\mathrm{mod} \; 8)$, the resulting code $T_1$ will have size $n \equiv 7 \; (\mathrm{mod} \; 8)$.  The next pair of doubly-even codes will require a number of qubits that is $2 \; (\mathrm{mod} \; 8)$, and so forth, so the size of the resulting triply-even code satisfies Eqn.~(\ref{eqn::code_size_constraints}) for any value of $t$.  Furthermore, transversal $X$ supported only on code $D_t^{(a)}$ commutes with the fusion process denoted by $\underset{ZZ}{\Longleftrightarrow}$, so it is a form of the logical $X_L$ operator for $T_t$.  Combined with the $X_L^{(T_{t-1})} X_L^{(b)}$ stabilizer from the logical Bell state, we see that $X_L = X^{\otimes n}$ is another form of this operator for $T_t$.  

\emph{Proof that Chaining Yields a Triply-Even Code} --- Using the procedure of the main text, $T_t$ is in fact a triply-even code, assuming the requirements given in Section~\ref{sec::general_construction} are satisfied.  Our proof is inductive: we assume that $T_{t-1}$ is triply-even, meaning that all of its $X$ stabilizers have weight $w(s^x) \equiv 0 \; (\mathrm{mod} \; 8)$, and they are not altered by conversion in Eqn.~(\ref{eqn::recursion}) since the type-B stabilizers from the logical Bell state necessarily commute with the stabilizers of $T_{t-1}$.  The $X$ stabilizers of each copy of $D_t$ all have weights that are multiples of four (as is necessary to be a doubly-even code), and the fusion process combines matching stabilizers from $D_t^{(a)}$ and $D_t^{(b)}$ into type-F $X$ stabilizers with weights being multiples of eight.  The final stabilizer to consider is the type-B stabilizer $X_L^{(T_{t-1})} X_L^{(b)}$.  All forms of logical $X$ operator $X_L^{(T_{t-1})}$ for $T_{t-1}$ have $w(X_L^{(T_{t-1})}) \equiv m \; (\mathrm{mod} \; 8)$, where $m = \pm 1$ depending on $t$ as in Eqn.~(\ref{eqn::code_size_constraints}).  The logical $X$ operator on $D_t^{(b)}$ after fusion is transversal $X$ supported on this block, which has weight $w(X_L^{(b)}) \equiv -m \; (\mathrm{mod} \; 8)$.  As a result, the type-B stabilizer  has weight $w(X_L^{(T_{t-1})} X_L^{(b)}) \equiv 0 \; (\mathrm{mod} \; 8)$.  Finally, the intersection of any of the aforementioned stabilizer generators has weight that is a multiple of four, which preserves weight modulo eight for any product of generators.  Consequently, the construction in Eqn.~(\ref{eqn::recursion}) ensures that all of the $X$ stabilizers of $T_d$ have weight that is a multiple of eight, and the code is triply-even.  Combined with the conditions we proved above and in Section~\ref{sec::general_construction} for code properties, transversal $T^{\otimes n}$ is the logical operator $T$ or $T^{\dag}$, depending on $t$.  Using induction to extend these results from the trivial case $T_0$ to any order, we have proved that our procedure constructs triply-even codes of increasing size with transversal $T$ gate.

\emph{Proof of Triply-Even Code Distance} --- The distance of $T_t$ is $d = 2t+1$, which we prove inductively by explicitly enumerating the logical operators.  Using the recursion relation in Eqn.~(\ref{eqn::recursion}), there are only two forms of the minimum-weight logical $Z_L^{(T_t)}$ operator: one comes from a $Z_L$ in $T_{t-1}$ and the other from $Z_L$ in $D_t$, and both have weight $d$ or higher after conversion to a triply-even code.  For the logical operator inherited from $T_{t-1}$, we presume that any logical operator $Z_L^{(T_{t-1})}$ has minimum weight $d-2$ (this will be true by induction starting from $T_0$).  This operator alone does not commute with the type-B stabilizer $X_L^{(T_{t-1})} X_L^{(b)}$, so this form of $Z_L^{(T_t)}$ requires at least two more physical $Z$ operators in the fused $D_t$ bilayer, located in the same horizontal position in these two codes (a ``vertical pair''), thereby summing to weight $d$ or higher.  Note that any number of vertical pairs will commute with the type-F stabilizers.  There is a degeneracy in the number of weight-$d$ logical operators of this form equal to the number of qubits in $D_t$ times the number of configurations for $Z_L^{(T_{t-1})}$.  The other form of logical $Z_L^{(T_t)}$ is inherited from $D_t$, where each physical $Z$ is located on a vertical edge in the bilayer at the same horizontal position as in $Z_L^{(D_t)}$.  When we say a physical $Z$ operator is ``located on a vertical edge,'' it is on precisely one of the two qubits in the bilayer touching that edge, but it could be on either, subject to overall parity constraints given below.  The total number of physical $Z$ operators located in the $D_t$ bilayer is at least $d$ (the distance of $D_t$), and the number of physical $Z$'s in the $D_t^{(b)}$ layer must be even for the operator to commute with the type-B stabilizer.  The degeneracy of this form at minimum weight is $2^{2t}$ times the number of weight-$d$ operators in $D_t$.  Any logical $Z_L^{(T_t)}$ with no $Z$'s in the $T_{t-1}$ block must be explicitly of this form or a combination with vertical $Z$ pairs (while maintaining even parity in $D_t^{(b)}$ layer), because any even number of vertical $Z$ pairs is a gauge operator from the fusion process.  These are the only two types of $Z$ configurations that commute with the $X$ stabilizers but are not in the $Z$ stabilizer group.

The enumeration above covers all logical $Z_L$ operators of $T_t$ of minimum weight $d$.  There are no $Z$ operators of lower weight (only stabilizers), so the code has distance $d$ to $Z$ errors.  Every $X$ stabilizer has a corresponding $Z$ stabilizer in the same configuration, so the code also has distance at least $d$ to $X$ errors as well, which completes the proof: our construction produces a triply-even code $T_t$ with distance $d$.  In fact, when $T_t$ is operated as a fixed code with no gauge qubits (not as a subsystem code), it has many more $Z$ stabilizers and a distance to $X$ errors that is significantly higher than $d$.  It can be shown that for our construction using [4.8.8] color codes, the $X$ distance is $2(t+1)^2 - 1$, which we conjecture to hold in general.  However, we do not delve further into the matter as it does not directly impact our proposal for subsystem codes (where the $X$ distance is only $d$ due to gauge fixing).

\emph{Arguments for No Threshold in Chained Triply-Even Codes} --- We outline here an argument that chained triply-even codes, including the 2D gauge color codes developed in the main text, do not have an error threshold to $Z$ errors that are independent and identically distributed (IID).  The fundamental weakness of all chained triply-even codes is the that type-B stabilizers (the high-weight stabilizers from logical Bell states) are increasing in weight with code order and have insufficient redundancy to uniquely identify physical errors.  To follow the arguments below, the reader should be familiar with the preceding properties of triply-even codes.  At a summary level, the proof is as follows.  We describe how to perform maximum-likelihood decoding of the error syndrome by incorporating the subset of the syndrome supported on bilayers incrementally from lowest order.  In the limit where code order goes to infinity, the type-B stabilizers that connect bilayers will be unable to distinguish complementary error patterns.  As a result, the logical error probability will not decrease exponentially with code distance.  

For simplicity, we consider only $Z$ errors (physical and logical).  The lack of a threshold to $Z$ errors implies no threshold in any realistic error model where there is necessarily a nonzero probability of $Z$ error.  Let us define an ``error coset'' and its associated probability.  The $Z$ stabilizer group $\mathcal{S}^z$ is the group of all elements of the code stabilizer that are tensor products of just physical $Z$ and identity operators; for the present analysis, this group also includes all $Z$ gauge operators.  For a configuration of physical $Z$ errors $e^z$, the error coset $e^z \mathcal{S}^z$ is the set formed by the product of $e^z$ and each element of $\mathcal{S}^z$.  The notion of an error coset is important because each element of the error coset is a distinct error configuration, but all such configurations will generate the same syndrome.  Moreover, there is a second error coset of interest, $e^z Z_L \mathcal{S}^z$, which also has the same syndrome but differs by logical Z from the elements of the first coset.  As a result, a given syndrome could correspond to any element of the two error cosets.  Moreover, all stabilizers have even weight, and the logical operator has odd weight, so all elements of one coset will have even weight, while the other will have odd weight.  The prior probability of seeing the syndrome for a given error coset, assuming IID physical $Z$ errors with probability $p$, is
\begin{equation}
\mathrm{Pr}\left(e^z \mathcal{S}^z\right) = \sum_{h \in e^z\mathcal{S}^z} p^{w(h)} (1-p)^{n - w(h)},
\end{equation}
where $n$ is the total number of qubits and $w(h)$ is the Pauli weight of operator $h$ (see Section~\ref{sec::general_construction}).  In maximum-likelihood decoding~\cite{Bravyi2014}, the syndrome is used to identify a consistent error pattern $e^z$ and the probabilities for the cosets $e^z \mathcal{S}^z$ and $e^z Z_L \mathcal{S}^z$.  The corrective action is to apply $e^z$ or $e^z Z_L$ to the physical qubits, depending which belongs to the more probable coset; the logical error probability is the renormalized probability of the other coset.

As before, we establish our maximum-likelihood decoder for $T_t$ using a recursive construction, then apply it inductively.  Note also that we use the maximum-likelihood decoder just for our argument, without any claim that it would be efficient to compute.  A chained triply-even code is constructed using the recursive construction in the main text, repeated here for convenience:
\begin{equation}
\overbrace{T_{t-1} \vert D_t^{(b)}}^{\mathrm{Bell}} \underset{ZZ}{\Longleftrightarrow} D_t^{(a)} \rightarrow T_t.
\label{eqn::recursive_appendix}
\end{equation}
The $Z$ stabilizer group $\mathcal{S}^z$ that is supported on the fused bilayer of $D_t^{(a)}$ and $D_t^{(b)}$ consists of the $Z$ stabilizers in those doubly-even codes and the new gauge operators measured for fusion.  Ignoring for the moment the type-B stabilizers that connect to $T_{t-1}$, this bilayer code has two logical qubits, so there are four error cosets.  Let $Z_L^{(a)}$ and $Z_L^{(b)}$ be the logical $Z$ operators on $D_t^{(a)}$ and $D_t^{(b)}$, respectively.  For a given syndrome in the type-F $X$ stabilizers supported on the $D_t$ bilayer, let $e^z$ be an error configuration that has an even number of errors in block (a) and no errors in block (b); such a configuration always exists, with no loss of generality.  The probabilities of the four error cosets in the $D_t$ bilayer are
\begin{align}
\lambda_{++} & = & & \mathrm{Pr}\left(e^z \mathcal{S}^z\right) & \\
\lambda_{-+} & = & & \mathrm{Pr}\left(e^z Z_L^{(a)} \mathcal{S}^z\right) & \\
\lambda_{+-} & = & & \mathrm{Pr}\left(e^z Z_L^{(b)} \mathcal{S}^z\right) & \\
\lambda_{--} & = & & \mathrm{Pr}\left(e^z Z_L^{(a)} Z_L^{(b)} \mathcal{S}^z\right). & 
\end{align}
The subscripts have an intuitive meaning: $++$ corresponds to error configurations with even weight on blocks (a) and (b), $-+$ corresponds to odd weight on (a) and even weight on (b), \emph{etc}.  Furthermore, we presume that the two error-coset probabilities are known for the stabilizer group supported on $T_{t-1}$ (these will be constructed inductively).  Label these probabilities $P_+^{(t-1)}$ and $P_-^{(t-1)}$ for the even-weight and odd-weight cosets, respectively.

Before completing the decoding procedure, we note some important symmetries in the error cosets.  For a given doubly-even bilayer, the four cosets can be grouped into two pairs, where the cosets in a pair differ only by multiplication with two $Z$ errors along a vertical edge, which is one form of $Z_L^{(a)} Z_L^{(b)}$ in the bilayer code.  For example, every element of $e^z Z_L^{(a)} \mathcal{S}^z$ can be transformed into an element of $e^z Z_L^{(b)} \mathcal{S}^z$ simply by multiplying by $Z \otimes Z$ along a vertical edge that currently has just one $Z$ error; the resulting product simply moves the $Z$ error from top to bottom (or vice versa) in the bilayer, which does not change the number of errors.  The probability of these two error configurations are equal assuming IID errors, so the sums over all such configurations are also equal ($\lambda_{-+} = \lambda{+-}$ always), for any physical error rate.  For the same reason, any element of $e^z \mathcal{S}^z$ has a corresponding element in $e^z Z_L^{(a)} Z_L^{(b)} \mathcal{S}^z$ by flipping a $Z$ location whenever there is at least one vertical edge that has exactly one error.  The only time this condition fails is if $e^z$ is identity, which corresponds to the ``null'' syndrome of having +1 eigenvalue for all type-F stabilizers in the bilayer.  As a result, $\lambda_{++} = \lambda{--}$, except for the null syndrome, which will be an important special case.

To complete the maximum-likelihood decoding, ``turn on'' the type-B stabilizers.  Only products of error cosets consistent with the type-B $X_L^{(T_{t-1})} X_L^{(b)}$ stabilizers will remain, and the type-B $Z_L^{(T_{t-1})} Z_L^{(b)}$ stabilizer will cause error cosets to merge (their probabilities sum because they were disjoint sets).  Without loss of generality, we assume that the type-B $X$ stabilizer is even parity (+1 eigenvalue), so the error-coset probabilities for $T_t$ will be given by
\begin{equation}
\left[ \begin{array}{c} \tilde{P}_+^{(t)} \\ \tilde{P}_-^{(t)} \end{array} \right] = \left[ \begin{array}{cc} \lambda_{++} & \lambda_{+-} \\ \lambda_{-+} & \lambda_{--} \end{array} \right] \left[ \begin{array}{c} P_+^{(t-1)} \\ P_-^{(t-1)} \end{array} \right],
\end{equation}
where the tildes on LHS indicate that these probabilities are not normalized.

The key insight to the proof is the following.  When there is a non-null syndrome in the $D_t$ bilayer, the bilayer coset probabilities will come in two pairs of equal values: $\lambda_{++} = \lambda_{--}$ and $\lambda_{-+} = \lambda_{+-}$.  The first symmetry is broken only when there is a null syndrome matching of no errors, in which case generally $\lambda_{++} \neq \lambda_{--}$.  Due to this symmetry, whenever a bilayer detects an error, it contributes essentially nothing to reducing probability of logical error.  With increasing code distance, the larger doubly-even bilayers will detect an error almost surely (\emph{i.e.} syndrome is not null), so the logical error rate cannot be suppressed arbitrarily towards zero.  
    
For a non-null bilayer syndrome and +1 type-B stabilizer, the iteration for updating coset probabilities with normalization and symmetric values is
\begin{equation}
\left[ \begin{array}{c} P_+^{(t)} \\ P_-^{(t)} \end{array} \right] = \frac{1}{\lambda_{++} + \lambda_{+-}} \left[ \begin{array}{cc} \lambda_{++} & \lambda_{+-} \\ \lambda_{+-} & \lambda_{++} \end{array} \right] \left[ \begin{array}{c} P_+^{(t-1)} \\ P_-^{(t-1)} \end{array} \right].
\end{equation}
Including the normalization coefficient, the stochastic matrix on the RHS will pull probabilities towards the fixed point of [0.5,0.5], meaning that the non-null syndrome will actually lead to increasing logical error rate.  In practice, if the doubly-even code upon which the bilayer is based has a threshold to $Z$ errors (for example, color codes~\cite{Landahl2011,Stephens2014}), then the coset probabilities will be separated exponentially in code distance (in other words, $\lambda_{++} \gg \lambda_{+-}$ or $\lambda_{++} \ll \lambda_{+-}$), so the increase in logical error rate due to this effect will be negligible.  However, this symmetry is broken when the bilayer has a null syndrome, in which case the renormalized coset probabilities can change significantly.  From this we can infer a general rule of thumb: adding doubly-even bilayers to extend code distance will only suppress logical error probability if there is significant likelihood of a null syndrome in each bilayer.

To estimate the behavior of extending code order to infinity, we consider the expectation value of the logarithm of the logical error rate, or $E\left[ \log \left(\epsilon_L\right) \right]$.  For each increment in code order, the logical error rate is suppressed by some factor that is lower-bounded by $C_t \epsilon$ \emph{for a null syndrome in the $D_t$ bilayer}, where $\epsilon$ is the physical error rate and $C_t > 1$ is a coefficient that depends on order $t$.  When the $D_t$ syndrome is not null, the logical error rate actually increases (see argument above).  A simple lower bound is to say that logarithm of probability decreases by less than $\left\vert \log \epsilon \right\vert$ per unit increase in order.  As a result,
\begin{equation}
E\left[ \log (\epsilon_L) \right] \ge \sum_{t = 1}^\infty \mathrm{Pr}(\textrm{null syndrome at order t}) \log (\epsilon).
\end{equation}
The probability of a null syndrome will decrease faster than $1/t$, so the sum is finite, meaning that the expected value of the logarithm of logical error probability is finite.  Using Jensen's inequality,
\begin{equation}
\log\left( E[\epsilon_L] \right) \ge E \left[ \log (\epsilon_L) \right],
\end{equation}
so the logarithm of the expected logical error rate is bounded above negative infinity.  This of course means that the logical error rate is bounded above zero for $t \rightarrow \infty$, so the logical error rate cannot be suppressed to arbitrarily low values simply by increasing code order.  However, our intuition above suggests that the logarithm of the logical error rate can be suppressed by about $\left\vert \log \epsilon \right\vert$ for a null syndrome in each bilayer, which is likely when the number of qubits in the bilayer, which scales as $4t^2 + O(t)$, is less than $1/\epsilon$.  If we assume that increasing the order of the code will only suppress logical error effectively up to this point, then we set the maximum effective order at $t = O(1/\sqrt{\epsilon})$.  The expected value for logarithm of logical error rate is $O(\epsilon^{-1/2}\log \epsilon)$, so the logical error rate is roughly estimated to be $\exp \left[O(\epsilon^{-1/2}\log \epsilon) \right]$.  This nonlinear functional form still allows for logical error rate to be suppressed well below the physical rate, for sufficiently low values of $\epsilon$.  Because of the rough approximations invoked above, we hesitate to call this a proof that the chained triply-even codes have no threshold, but such a result would not be surprising.

\section{Algorithm for Decoding Error Syndromes in Chained Triply-Even Codes}
\label{app::decoder}
This appendix presents a ``minimum-weight'' algorithm for decoding the error syndrome of any chained triply-even code, not just color codes.  The procedure is efficient in the following way: it requires an oracle for decoding each doubly-even code used in the code conversion, and the computational complexity of the triply-even decoder is the sum of the costs for implementing oracles on each of the doubly-even codes, followed by a step to synthesize those results, where the cost is linear in $t$, the number of fused doubly-even bilayers (or equivalently, linear in the code distance $d$).  Furthermore, the decoder we describe is ``full distance,'' meaning it corrects any error of weight $t$ or less, if the doubly-even decoders are full distance.  The latter point can affect complexity, as currently known full-distance, minimum-weight decoders for triangular color codes do not run in polynomial time~\cite{Landahl2011,Stephens2014b,Delfosse2014}, though the tensor-network decoder in Ref.~\cite{Bravyi2014} is noted to be promising~\cite{Landahl2014}.  For the purposes below, we only presume a full-distance decoder for each doubly-even code is available.  We also assume perfect stabilizer measurement, and we leave the matter of handling faulty measurements to future work.

We focus on decoding $Z$ errors (using $X$ stabilizers), as this is all that is necessary for operating the universal color codes in the manner described in the main text.  The sequence we propose is to expand a doubly-even code to a triply-even code only to implement a logical $T$ gate, then collapse back to a doubly-even code by making transversal $X$-basis measurements on the rest of the qubits outside the support of the doubly-even code (see Section~\ref{sec::bilayer_color_codes}).  With this approach, only $Z$ errors can be detected by the triply-even code, but the final doubly-even code can detect $X$ errors as well using its own decoding procedure, as it has a sufficient set of persistent $Z$ stabilizers to correct any configuration of $t$ or fewer $X$ errors.  In combination, the two methods can correct any error up to weight $t$.  

The decoding procedure consists of two steps, because we separate the $X$ stabilizers into two types, type-F (``fusion'') for each fused pair of doubly-even codes and type-B (``Bell'') for the $X_L X_L$ stabilizer from the logical Bell pairs.  First, use the doubly-even oracle decoder to decode the syndrome for the set of type-F $X$ stabilizers in each fused bilayer of doubly-even $D_\mu$ codes, for $\mu = 1,\ldots,t$.  Treat the type-F stabilizers in the order-$\mu$ bilayer as if they were stabilizers in a single $D_\mu$ block; each type-F stabilizer is a fused pair of matching $X$ stabilizers from two copies of $D_\mu$, so there is a direct mapping.  The resulting minimum-weight error configuration also applies to the fused bilayer, but there is ambiguity as to which code block each $Z$ error is located in.  Specifically, each error in the identified configuration is associated with a vertical edge, but it is not yet determined whether the error is located in top or bottom layer.  Perform this procedure separately on the fused code blocks $D_1$, $D_2$, ... $D_t$, and record the minimum-weight error configurations, which are used by the next step.

\begin{figure}
	\centering
  \includegraphics[width=8.3cm]{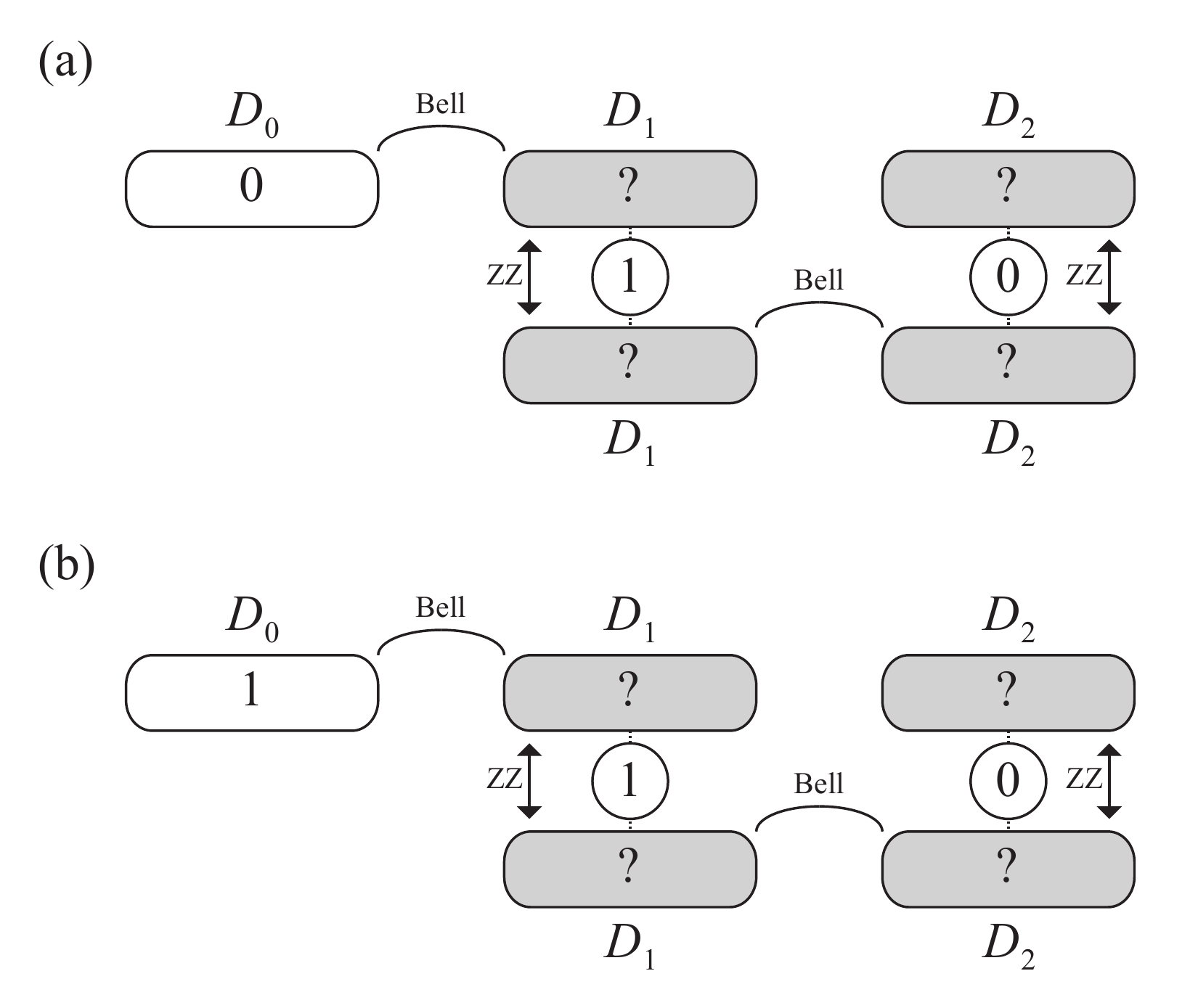}
  \caption{Example configuration of $Z$ errors to initialize the triply-even syndrome decoder.  The decomposition of the triply-even code into doubly-even codes is shown, with the connections through logical Bell states (``Bell'') and fusion (``ZZ'') labeled.  Note that this corresponds to $T_2$ in Fig.~\ref{fig::bilayer_structure} of the main text.  As this is a soft decoder, two configurations are considered with (a)~zero $Z$ errors on $D_0$ and (b)~one $Z$ error on $D_0$, as shown by the number in that box.  By processing the fused doubly-even codes, errors are located on vertical edges, but there is ambiguity as to whether the error should be in the top or bottom code (represented by grayed boxes with question marks).  The circled number between the two copies of the same code ($D_1$ or $D_2$ here) show an example configuration of located vertical-edge errors that could occur after processing the type-F stabilizer measurements as syndromes in a doubly-even code.}
  \label{decoder_init}
\end{figure}

The second step incorporates the remaining type-B stabilizers.  Use a ``soft'' decoder which constructs two possible error configurations and decides which has lower weight at the end of the decoding process.  Initialize the two configurations in the following way.  Taking the $Z$-error configurations associated with each fused doubly-even code pair (identified in the first step), assign these errors to ``vertical edges'' between the two blocks.  We have not yet assigned these errors to top or bottom block, and in fact only the odd/even parity in either block will be detectable, meaning many different error patterns are degenerate.  Figure~\ref{decoder_init} shows an example of the decoder state at this point.  For the two prospective configurations, assign no error to $D_0$ (Fig.~\ref{decoder_init}a) and one error to $D_0$ (Fig.~\ref{decoder_init}b).  

\begin{figure}
	\centering
  \includegraphics[width=8.3cm]{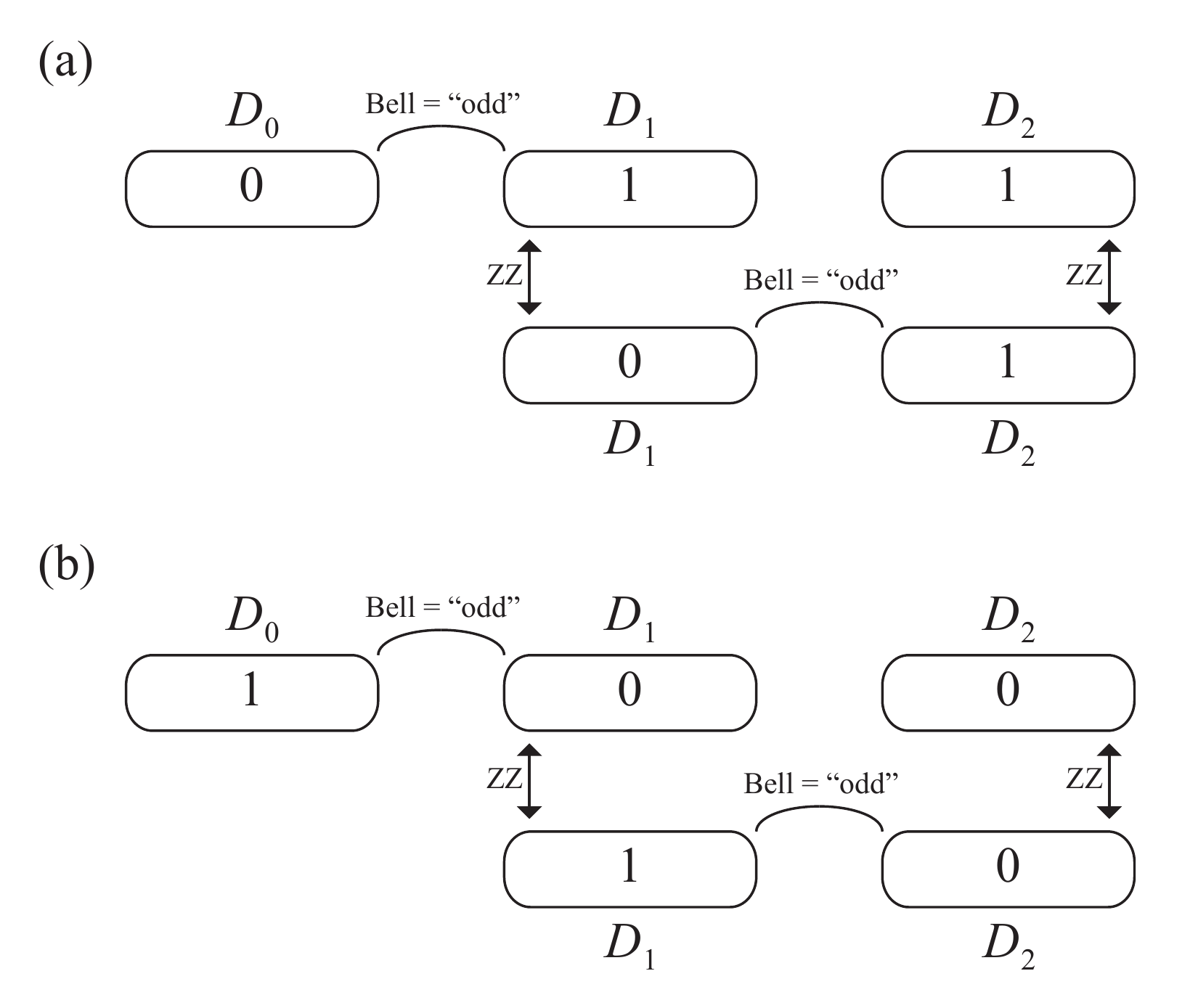}
  \caption{Final error configurations in the decoder after cascading through the type-B stabilizers.  For this example, both have been assigned -1 eigenvalue, which corresponds to odd parity of errors.  (a)~The ambiguous vertical edge in $D_1$ is assigned to the top code.  Since the bottom $D_1$ has even error parity while the type-B stabilizer is odd, the bottom $D_2$ must also have an odd number of errors.  To achieve this, a vertical pair is inserted, placing one error each in top and bottom $D_2$.  The total weight of this configuration is now three.  (b)~The ambiguous vertical edge is placed in the bottom $D_1$ block, because $D_0$ has an error and the first type-B stabilizer has odd parity, so top $D_1$ must be even.  Placing zero errors in top and bottom $D_2$ satisfies the second type-B stabilizer.  The weight of this configuration is two.  This example highlights the importance of the soft decoder, as sometimes the minimal-weight matching changes when vertical pairs must be inserted, as must be done in this example for case~(a).}
  \label{decoder_final}
\end{figure}

Next, in a cascade fashion, move rightward through the sequence to fix the parity of $Z$ errors in top or bottom code blocks.  For example, the first type-B stabilizer will fix the joint parity of errors in $D_0$ and the left-facing copy of $D_1$ (using labels for blocks of qubits before they were fused together, and ``facing'' direction refers to the connection formed by a type-B stabilizer in Fig.~\ref{decoder_init}).  The next type-B fixes the joint parity of right-facing $D_1$ and left-facing $D_2$, and so forth.  To satisfy the parity constraints, assign the necessary number of vertical-edge errors to the left-facing block spanned by the type-B stabilizer; the remaining errors are assigned to the other block in this fused pair, which combines with the next type-B stabilizer to fix parity in the next block to the right, and so forth.  One final modification is that one might need to assign an odd number of errors to the left-facing block in a fused pair when there are zero vertical-edge errors.  In this case, insert a pair of $Z$ errors that are vertically aligned, meaning in the same place in both top and bottom; any location works, which is another form of error degeneracy.  Figure~\ref{decoder_final} shows the final decoder state for the example in Fig.~\ref{decoder_init}; in particular, Fig.~\ref{decoder_final}a provides an example where the decoder inserts a vertical pair of errors.

After cascading through the type-B stabilizers, the parity combination of the two configurations being considered is $Z_L$ on the triply-even code, which is necessarily odd in weight, so one configuration is even while its complement is odd.  Finally, the decoder selects the minimum-weight error configuration, which is unambiguous.

\emph{Proof}---To prove that the syndrome decoder corrects any set of $Z$ errors up to weight $t$, we use the recursion for increasing the order of a triply-even code in Eqn.~(\ref{eqn::recursive_appendix}).  Let us suppose that the syndrome of $X$ stabilizers supported only within $T_{t-1}$ can be decoded in a manner that yields two configurations of $Z$ errors with the following properties: (1) their parity combination is a logical $Z_L$ operator (we say they are ``complementary''), which implies one is even in weight and the other odd; and (2) each configuration has the minimum weight of an even or odd error pattern that is consistent with the syndrome.  By definition of the code distance, the sum of their weights must be at least $2t-1$, the minimum weight of a logical operator.  These properties hold trivially for $T_0$ by considering an error or not on the single qubit, and at higher orders they will hold by induction.  As a result, the minimum-weight, complementary error configurations for $T_{t-1}$ are known.  We further assume that one can decode the doubly-even code $D_t$ to determine the minimum-weight error configuration for any syndrome, but knowing the complementary configuration is not necessary.

The following procedure is useful for identifying a minimum-weight error pattern.  When the syndrome decoder is applied to the recursion above, the stabilizers of the $D_t$ bilayer can always locate up to $t$ vertical-edge errors, because we assume the decoder for $D_t$ can locate any set of errors up to  weight $t$.  A vertical pair is two errors on the same vertical edge, and any two vertical pairs is a $Z$ stabilizer.  To be clear on terms, a ``vertical-edge error'' is a single physical $Z$ operator, with ambiguity as whether it is in top or bottom layer; a ``vertical pair'' is two $Z$ operators on both qubits connected by a vertical edge.  A single vertical pair commutes with type-F stabilizers but not any overlapping type-B stabilizer, which is why the decoder inserts a vertical pair into a prospective error configuration when a type-B stabilizer detects error but no vertical edges are present.  If there are greater than zero vertical-edge errors, the effect of any number of vertical error pairs on the stabilizer is equivalent to having no vertical pairs and perhaps modifying a vertical-edge error location (switching top for bottom), where the reduction uses parity combination with $Z$-type fusion stabilizers.  As a result, we need only consider the number of vertical-edge errors (\emph{i.e.} vertical edges that must have one error), except for one case.  If there are no vertical-edge errors identified for the $D_t$ bilayer, then the decoder will assign either zero or one vertical pair of $Z$ errors in the bilayer, as needed to satisfy parity in type-B stabilizers.  In all cases, the number of vertical-edge errors identified by the $D_t$ decoder is a lower bound for the number of errors located in the $D_t$ bilayer for any minimum-weight error configuration, which is useful for showing that a prospective error configuration has minimum weight.

The decoder for $T_{t-1}$ identifies the two minimum-weight, complementary error configurations within that code block.  The decoder for $D_t$ identifies the minimum number of vertical-edge errors.  If that number is greater than zero, then the total number of errors is correct for both complementary configurations, and the type B stabilizer between $T_{t-1}$ and $D_t^{(b)}$ will fix the parity of errors at top or bottom of vertical edges.  If the number of vertical-edge errors in the $D_t$ bilayer is zero, then the type-B stabilizer will force one of the configurations to insert a vertical pair of errors, simply due to parity.  This pair is the minimum number of errors to insert, because a single error in the $D_t$ bilayer would be detected by the type-F stabilizers.  The combination of the two complete error configurations at the output of the decoder is a logical $Z_L$ operator for $T_t$, so they are complementary.  A simple proof of this fact is that their parity combination is a logical $Z$ in $T_{t-1}$ (by assumption) and an odd number of vertical pairs in the $D_t$ bilayer (as required by the type-B stabilizer).  This operator can be reduced through the stabilizer and gauge groups to $Z_L^{(T_{t-1})}$ and one vertical pair in $D_t$, which is $Z_L^{(T_t)}$, as shown in Appendix~\ref{app::properties}.  This completes the recursion required for the proof to apply for any order of chained triply-even code using induction.  The two error configurations are minimum-weight and complementary (their combined weight is $2t+1$ or higher), so this decoder can always correct $t$ or fewer errors.

\bibliography{color_codes}

\end{document}